\documentclass[aps,twocolumn,superscriptaddress,floatfix,longbibliography]{revtex4-2}
\usepackage{dcolumn}
\usepackage{amsmath}
\usepackage{amssymb}
\usepackage{amsthm}
\usepackage{graphicx}
\usepackage{bm}
\usepackage[T1]{fontenc}
\usepackage{color}
\usepackage{url}
\usepackage[bf]{subfigure}
\usepackage{rotating}
\usepackage{scalefnt}
\usepackage{multirow}

\usepackage{scalefnt}
\usepackage{times}
\usepackage{mathtools}
\usepackage{bbm}

\usepackage{float}
\usepackage{enumerate}

\usepackage{nicefrac}

\newcommand{\ContThreshold}{\Theta_{\lambda}}

\newcommand{\AnniThreshold}{\Theta_{\alpha}}

\newcommand{\Nsim}{n_{\text{sim}}}

\newcommand{\OrderParam}{z} 

\newcommand{\Susceptibility}{\chi}

\begin{document}

\title{Rumor propagation on hypergraphs}

\author{Kleber A. Oliveira}
\altaffiliation{These two authors contributed equally.}
\affiliation{Department of Mathematics, Munster Technological University, Cork, Ireland}

\author{Pietro Traversa}
\altaffiliation{These two authors contributed equally.}
\affiliation{Institute for Biocomputation and Physics of Complex Systems (BIFI), University of Zaragoza, Zaragoza 50009, Spain}
\affiliation{Department of Theoretical Physics, Faculty of Sciences, University of Zaragoza, Zaragoza 50009, Spain}

\author{Guilherme Ferraz de Arruda}
\affiliation{Institute of Physics Gleb Wataghin, University of Campinas (UNICAMP), Campinas, SP, Brazil}

\author{Yamir Moreno}
\affiliation{Institute for Biocomputation and Physics of Complex Systems (BIFI), University of Zaragoza, Zaragoza 50009, Spain}
\affiliation{Department of Theoretical Physics, Faculty of Sciences, University of Zaragoza, Zaragoza 50009, Spain}

\begin{abstract}
The rapid spread of information and rumors through social media platforms, especially in group settings, motivates the need for more sophisticated models of rumor propagation. Traditional pairwise models do not account for group interactions, a limitation that we address by proposing a higher-order rumor model based on hypergraphs. Our model incorporates a group-based annihilation mechanism, where a spreader becomes a stifler when the fraction of hyperedges aware of the rumor exceeds a threshold. Our model has two distinct subcritical behaviors: exponential and power-law decay, which can coexist depending on the heterogeneity of the hypergraph. Interestingly, our analysis reveals continuous phase transitions in both homogeneous and heterogeneous hypergraphs, challenging the idea that higher-order interactions lead to discontinuous transitions. Finally, we validate our model using empirical data from Telegram and email cascades, providing additional evidence that real-world rumor propagation tends to occur near criticality. These results open the door to a more detailed understanding of rumor dynamics in higher-order systems.
\end{abstract}

\maketitle

\section{Introduction}

Rumor and information propagation pose a challenge that increasingly demands systematic understanding as human communication scales up in volume and complexity, affecting various societal aspects. In particular, a significant part of this communication takes place in social media platforms, one of the main sources of contemporary information consumption~\cite{allen2020fakenews}. As public referenda and elections require informed decisions from individuals and they turn to such platforms, they are targeted as fertile ground for misinformation campaigns and ideological polarization phenomena~\cite{lorenzspreen2023democracy}. Adequate policies to regulate platforms need to be built from scientific knowledge, drawing from the mathematical modeling of mechanisms explaining the emergent collective behavior~\cite{zhang2016information}. Although much of the empirical research on the spread of misinformation has been conducted on microblogging platforms~\cite{aimeur2023misinfo, meel2020rumor}, mobile messenger apps are widely used in populous countries such as India, Brazil, Nigeria, and Indonesia~\cite{bradshaw2019global}. As a result, much of the understanding gained from network diffusion models may not be directly applicable to mobile messenger apps such as WhatsApp or Telegram. In these platforms, users are organized into groups, and the information is typically forwarded from group to group. Dyadic interactions are no longer sufficient to adequately describe the spread of group-mediated rumors, which means that mobile messenger apps are suitable objects for studying nonlinear dynamics through higher-order interactions~\cite{battiston2021higherorder}.

Spreading processes on higher-order interactions constitute one of the latest research frontiers in network science~\cite{ferrazdearruda2024contagion}. Compartmental models are featured in recent developments for both simplicial complex~\cite{iacopini2019simplicial,dearruda2020social,higham2022meanfield,FerrazdeArruda2023multistability,kiss2023simplex} and hypergraph~\cite{jhun2021epidemic,stonge2022influential,mancastroppa2023hypercore, kim2024higherorder}. For the latter, it has been recently proposed that the mechanism of changing social norms via critical mass~\cite{centola2018tipping} is behind rich phenomena on hypergraph contagion, such as multistability and hybrid phase transitions~\cite{dearruda2020social,higham2022meanfield,FerrazdeArruda2023multistability}. In this class of models, once a fraction of individuals in a group is infected, so are the remaining susceptible individuals in that same group. However, implementing a rumor-propagation annihilation process from the same principle remains unclear since to follow Daley -- Kendall's original proposal~\cite{daley1964epidemics}, which was further adapted in the equally important Maki -- Thompson model~\cite{maki1973mathematical}, the annihilation is to be produced from social interaction.

We introduce a higher-order rumor model that follows threshold-based propagation, as in~\cite{dearruda2020social,higham2022meanfield,FerrazdeArruda2023multistability}, and incorporates group-based annihilation, whose definition is inspired by how rumors spread in messaging applications such as Telegram. Our model builds on the fundamental principle that rumor annihilation occurs through interaction with individuals who are already familiar with the rumor, rather than being spontaneous, following the original proposal of the Daley-Kendall model~\cite{daley1964epidemics}, but extending this idea to a group-structured population. We found a rich behavior depending on the propagation and annihilation thresholds, including regions of exponential or power-law decay to an absorbing state in the subcritical regime. The boundaries of these regions can be non-trivial in the presence of heterogeneity, including regions where both behaviors can coexist, depending on the initial condition. Importantly, we find empirical evidence for our model using the Telegram public channels~\cite{baumgartner2020pushshift} dataset. At the same time, our results may also provide a mechanistic explanation for the observation that real-world rumors occur near criticality, as proposed in~\cite{notarmuzi2022universality}. The paper is organized as follows: in Sec.~\ref{Sec:The model}, we introduce the model. In Sec.~\ref{Sec:numerical}, we analyze its parameters, inspect its phase space as the parameters change, and also perform a finite-size simulation to characterize the phase transition. In Sec.~\ref{Sec:Interevent}, we bridge simulations and real-world scenarios by studying cascades on Telegram. A discussion is presented in Sec.~\ref{Sec:discussion}.
\section{The model}
\label{Sec:The model}
First, we define the hypergraph to represent group interactions in the rumor model. A hypergraph is a generalization of a graph in which edges connect not just two nodes but any number of nodes. These groups of nodes are called hyperedges, and the size of a hyperedge is called cardinality. Let $\mathcal{H} = \{ \mathcal{V}, \mathcal{E}\}$ be a connected hypergraph such that $\mathcal{V} = \{v_1, v_2, \cdots, v_N\}$ is the set of nodes, while $\mathcal{E} = \{e_1, e_2, \cdots, e_M \}$ is the set of hyperedges, where $e_j$ is a subset of $\mathcal{V}$. To simplify the notation, we will always use $i$ to index nodes and $j$ to index hyperedges. The hypergraph structure is completely described by an incidence matrix $\mathcal{I}$, defined as
\begin{equation}
    \mathcal{I}_{ij} =
    \begin{cases}
        1 \hspace{5mm} \text{if} \hspace{2mm} v_i \in e_j, \\
        0 \hspace{5mm} \text{otherwise.}
    \end{cases}
\end{equation}
It is also useful to define the degree of a node, $k_i= \sum_j \mathcal{I}_{ij}$, which corresponds to how many hyperedges it belongs to.

In a rumor model, node $v_i$ can be one of three states: (i) ignorant ($X_i = 1$), individuals who are not aware of the rumor, spreader ($Y_i = 1$), individuals who are aware of the rumor and spread it, and (iii) stifler ($Z_i = 1$), individuals who are aware of the rumor but no longer spread it. We track their states with three associated Bernoulli random variables $X_i (t)$, $Y_i (t)$, and $Z_i (t)$. The three random variables are linearly dependent and must satisfy the condition $X_i (t) + Y_i (t) + Z_i (t) = 1$. In addition, nodes change their states based on two types of processes: (i) the contagion process ($X \rightarrow Y$), in which ignorant nodes become spreaders, and (ii) an annihilation process ($Y \rightarrow Z$), in which spreaders become stiflers.

We model the contagion process as a higher-order contagion process with threshold dynamics, similar to~\cite{FerrazdeArruda2023multistability}. The contagion is a Poisson process that occurs at a rate $\lambda$ when the number of spreaders in a hyperedge reaches a certain threshold $\ContThreshold^{(j)}$. When this condition is met, all the ignorants in the hyperedge become spreaders, and we say that the hyperedge is activated. Mathematically, it is useful to introduce a random variable $T_j$ that counts the number of spreaders in a hyperedge $e_j$ such that
\begin{equation}
    T_j = \sum_{i:v_i \in e_j}Y_i = [\mathcal{I}_{\cdot j}]^T Y ,
\end{equation}
where $[\mathcal{I}_{\cdot j}]$ is the $j$-th column of $\mathcal{I}$ and $Y$ is the vector of Bernoulli variables $Y_i$ of size $N$. For each $e_j$, we compare $T_j$ with $\ContThreshold^{(j)}$ to determine if the hyperedge can be activated.

Complementarily, we model the annihilation process as a node-based process with a threshold defined on the degree of the node. Let the annihilation be a Poisson process that occurs when the number of activated hyperedges $v_i$ belongs to is above a certain threshold $\AnniThreshold^{(i)}$. We can define a vector $B(t)$ indexed by $j$ such that it is $1$ if $e_j$ has already been activated by contagion at time $t$ and $0$ otherwise. Then we write the number of activated hyperedges to which $v_i$ belongs as
\begin{equation}
    S_i = \sum_{j: v_i \in e_j}\prod_{\ell:v_\ell \in e_j}\left(1 - X_\ell\right) = B^T[\mathcal{I}_{i \cdot}]^T,
\end{equation}
where the summation runs over all hyperedges to which node $v_i$ belongs, $X_\ell$ tracks which neighboring nodes of $i$ are ignorant, and $[\mathcal{I}_{i \cdot}]$ is the $i$-th row of the incidence matrix. Note that, by definition, an activated hyperedge cannot contain ignorant nodes.

Thus, under these assumptions, the equation describing the time evolution of the probability of node $v_i$ being a spreader is given by
\begin{equation}\label{eq:model_dy}
    \frac{d \langle Y_i \rangle}{dt} = \left\langle \lambda X_i \sum\limits_{j: v_i \in e_j} H(T_j - \ContThreshold^{(j)}) - \alpha Y_i  H(S_i - \AnniThreshold^{(i)}) \right\rangle,
\end{equation}
where $\langle \cdot \rangle$ is the expectation operator, $H(x)$ is the Heaviside function which is zero for $x<0$ and 1 for $x\geq 0$, and the terms $\ContThreshold^{(j)}$ and $\AnniThreshold^{(i)}$ can depend on the hyperedge $e_j$ or the node $v_i$. In particular, we focus our study on the case where contagion is controlled by $\ContThreshold^{(j)} = \lfloor(\ContThreshold \times |e_j|) \rceil$ and annihilation is controlled by $\AnniThreshold^{(i)} =\lfloor(\AnniThreshold \times k_i)\rceil$, where $k_i$ is the degree of the node $v_i$. In this way, $\ContThreshold$ and $\AnniThreshold$ are numbers between $0$ and $1$ that can be fixed independently of hyperedge cardinalities and degrees. The function $\lfloor x \rceil$ rounds $x$ to its nearest integer. This choice is motivated by the intuition that the larger a group is, the harder it is to inform all group members. Similarly, for annihilation, the more groups one participates in, the more reluctant the spreader is to become a stifler.

\begin{figure*}[t]
    \centering
    \includegraphics[width=\linewidth]{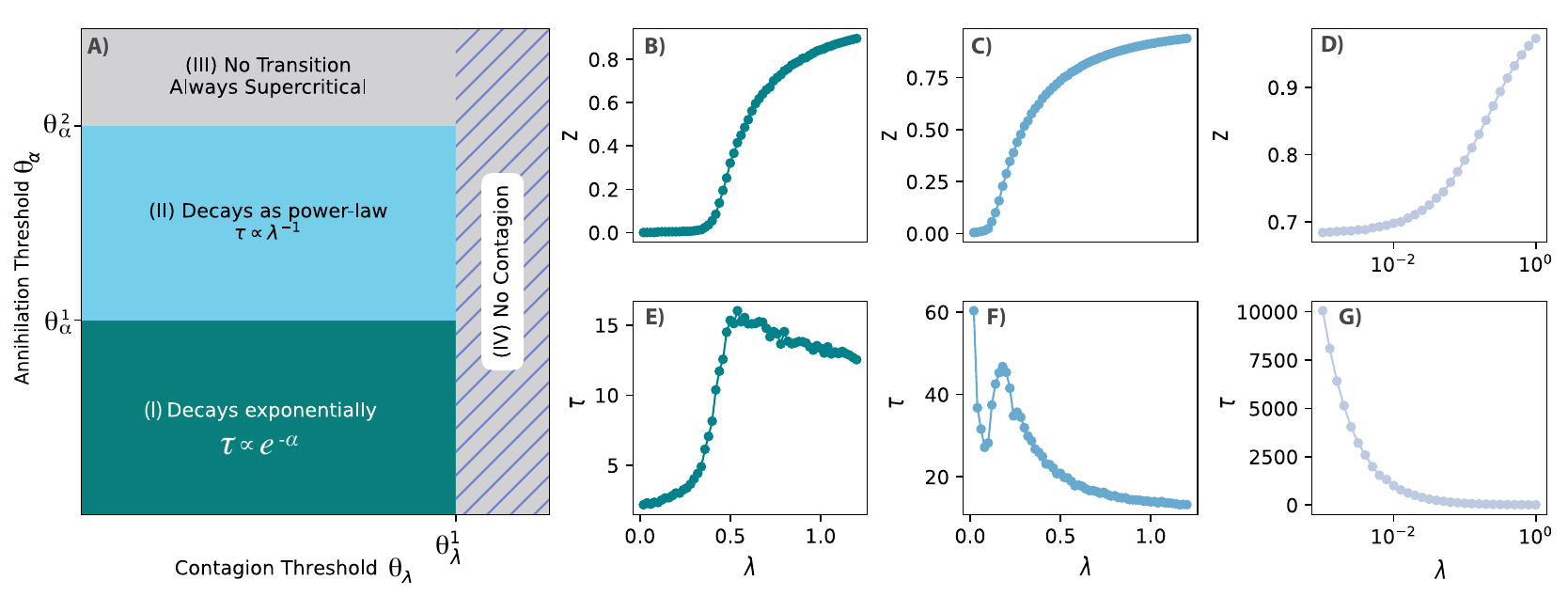}
    \caption{\textbf{Phase diagram of the hypergraph rumor model.} A) Schematic representation of the critical and subcritical behavior depending on the annihilation and contagion threshold parameters. If the contagion threshold is too high, no contagion is possible. If the annihilation threshold is low, each active node can recover exponentially fast, and the time to the absorbing state also decays exponentially with rate $\alpha$. As the annihilation threshold increases beyond a certain value, some spreading processes must occur to trigger the recovery processes, and thus, the absorbing time decreases as a power law as $\lambda^{-1}$. B) and E) show the order parameter and the absorption time in the exponential regime, respectively. This example was generated with $\AnniThreshold = 0.09$ and $\ContThreshold = 0.3$. Panels C) and F) show the order parameter and the absorption time in the power-law regime, respectively, with $\AnniThreshold = 0.4$ and $\ContThreshold = 0.35$. Finally, D) and G) show the order parameter and absorbing time in the case where the rumor always reaches a fraction of the population, so there is no transition. For this regime, we chose $\AnniThreshold = 0.5$ and $\ContThreshold = 0.2$. All examples are run on the same hypergraph with truncated Poisson-distributed cardinalities and degrees, i.e., $P(k) = \frac{1}{1 - e^{-\nu} - \lambda e^{-\nu}} \frac{\nu^k e^{-\nu}}{k!}$, for $k \geq 2$ and $\nu = 2$ (similarly for $|e|$), leading to an average of approximately $2.9$.}
    \label{fig:diagram}
\end{figure*}

To analyze our model, we focus on the time to reach an absorbing state, $\tau$, and the fraction of individuals reached by the rumor, $\OrderParam = \langle Z \rangle$. We rely on Monte Carlo simulations performed by the Gillespie algorithm described in Sec.~\ref{Sec:methods}. Both $\tau$ and $z$ are estimated using the average of $\Nsim$ simulations, which can vary depending on the experiment. Fig.~\ref{fig:diagram} illustrates the rich behavior observed in terms of regimes for different combinations of the two thresholds, where we can identify four regions:
\begin{enumerate}[(I)]
    \item A region whose subcritical regime decays exponentially into an absorbing state, i.e., $\tau \propto \exp (-\alpha)$, which is the same subcritical behavior as an SIS or a SIR on a graph~\cite{Mieghem2009,deArruda2028} (note that $\tau$ peaks at the critical point in figures~\ref{fig:diagram}E);

    \item A region whose subcritical regime decays as a power-law to an absorbing state, i.e., $\tau \propto \lambda^{-1}$, which is the same subcritical behavior as the Maki-Thompson model on a graph~\cite{ferrazdearruda2022subcritical} (note that $\tau$ is the largest at the smallest $\lambda$ in figures~\ref{fig:diagram}F);

    \item An active region where the rumor is always in the supercritical regime (note that for a non-zero $\lambda$, even when $\lambda \rightarrow 0$, $z>0$, and that a change in the derivative of $z$ is present, but there is no phase transition, in Fig.~\ref{fig:diagram}D);

    \item An inactive region where the rumor does not spread.
\end{enumerate}

The active and inactive regions are separated by $\ContThreshold^1$, below which the process is active, while above which the process cannot start because there are not enough spreaders to trigger the spreading. The three active regions are separated by the boundaries $\AnniThreshold^1$ and $\AnniThreshold^2$ (see Fig.~\ref{fig:diagram}A). The threshold $\AnniThreshold^1$ marks the regime where, below it, every node in the hypergraph recovers exponentially fast, provided the rumor is able to spread, while $\AnniThreshold^2$ is the value above which the rumor always reaches a non-zero fraction of the population, even for small but non-zero $\lambda$.

Two mechanisms explain the subcritical behavior in regions (I) and (II). In (I), the exponential behavior is due to a low annihilation threshold where the initial condition is sufficient to trigger the process. Thus, since the events are modeled by Poisson processes, the expected time for the next annihilation event is exponential with the parameter $\alpha$. On the other hand, in region (II), the power-law behavior is a consequence of a moderately high threshold and of $\alpha \gg \lambda$. In this case, we need at least one spreading event to reach the absorbing state, which takes, on average, $\lambda^{-1}$ time units. Note that these mechanisms are fundamental for describing the subcritical regime, but both coexist in the supercritical regime.

\begin{figure}
    \centering
    \includegraphics[width=\linewidth]{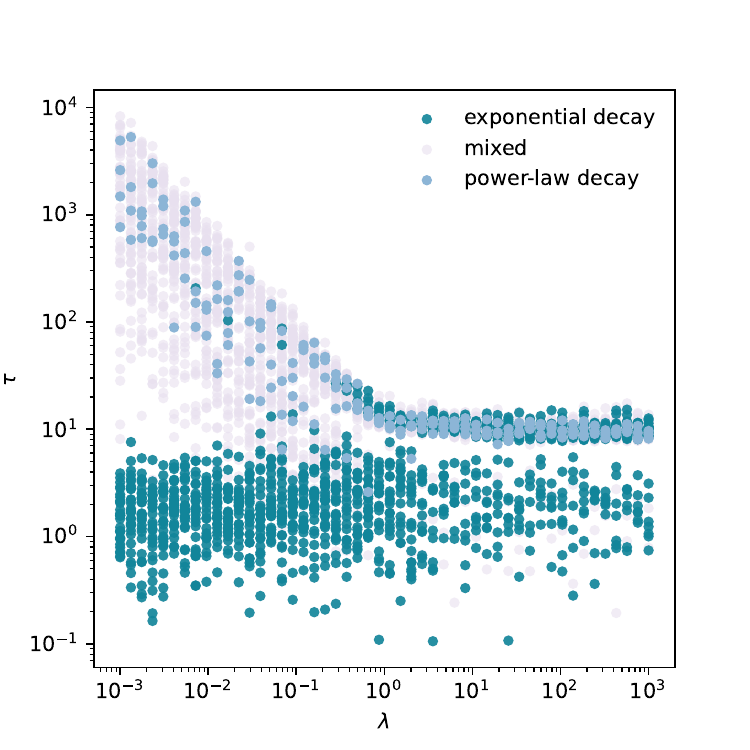}
    \caption{\textbf{Dependence of the subcritical regimes on the initial condition.} We show the absorption time $\tau$ as a function of $\lambda$ for different initial conditions. Simulations labeled as "exponential decay" have initial conditions satisfying $k_i \times \Theta_\alpha \leq 1$ $\forall i: v_i \in e_j$, where $e_j$ is the initially infected hyperedge. In the subcritical regime, these initial conditions necessarily lead to exponential decay. Instead, the initial conditions of the "power-law decay" simulations satisfy $k_i \times \Theta_\alpha > 1$ $\forall i: v_i \in e_j$. In these cases, a contagion process is forced to occur, leading to a power-law decay. The solutions labeled as ``mixed'' are those where we have different nodes satisfying different conditions. We considered a hypergraph with $N=10^4$ nodes, and both degree and cardinality follow power-law distributions as $P(k) \sim k^{-\gamma}$ and $P(|e|) \sim |e|^{-\gamma}$ with $\gamma = 2.3$, with $k_\text{min} = 2$ and $k_\text{max}= 100$. The simulation parameters are $\AnniThreshold = 0.1$, $\ContThreshold = 0.5$, and we run $\Nsim = 10^2$ independent simulations (initial conditions) for each value of $\lambda$. The critical point is around $\lambda_c \approx 0.12$.}
    \label{fig:AbsorbingTime_PL}
\end{figure}

Evaluating the bounds $\ContThreshold^1$, $\AnniThreshold^1$, and $\AnniThreshold^2$ is challenging, as they generally depend on the degree and cardinality distributions of the hypergraph. However, we can give some bounds independent of the distributions (see Sec.~\ref{Sec:methods} for their derivations), and we get
\begin{equation}\label{eq: Anni_bound}
    \AnniThreshold^1 = \frac{1}{k_{\text{max}}}, \; \AnniThreshold^2 = \frac{1}{k_{\text{min}}},
\end{equation}
where $k_{\text{min}}$ and $k_{\text{max}}$ are the minimum and maximum degrees, respectively. Similar considerations lead to the evaluation of the inactive region as
\begin{equation}\label{eq: Cont_bound}
    \ContThreshold^1 = 1 - \frac{1}{|e|_{\text{max}}}.
\end{equation}
We note that $\AnniThreshold^1$ and $\AnniThreshold^2$ define the extreme regions of this parameter space, i.e., (I) and (III), while the intermediate region can be a mixture of both behaviors, which can coexist for heterogeneous degree distributions. More specifically, since the rumor model has infinitely many absorbing states in the thermodynamic limit, in the subcritical regime, the process depends only on the initial conditions. In our simulations, we define the initial condition by randomly selecting a single hyperedge and setting all its nodes as spreaders.  In this case, these bounds can be interpreted in terms of the degrees of the nodes in this hyperedge. In this case, replacing $k_{\text{max}}$ and $k_{\text{min}}$ by $k_\text{max}^{(j)} = \max_{v_i \in e_j} k_i$ and $k_\text{min}^{(j)} =\min_{v_i \in e_j} k_i$ of the nodes in the initially activated hyperedge, $e_j$, in the previous expression~\eqref{eq: Anni_bound} will give us a new set of conditions. Fig.~\ref{fig:AbsorbingTime_PL} illustrates the dependence of the subcritical regimes on the initial condition of heterogeneous hypergraphs. Note, however, that region (II) does not exist for a uniform and regular hypergraph since both bounds are the same, but this region does exist for non-regular and non-uniform homogeneous systems.

\begin{figure*}[t]
    \centering
        \includegraphics[width=\linewidth]{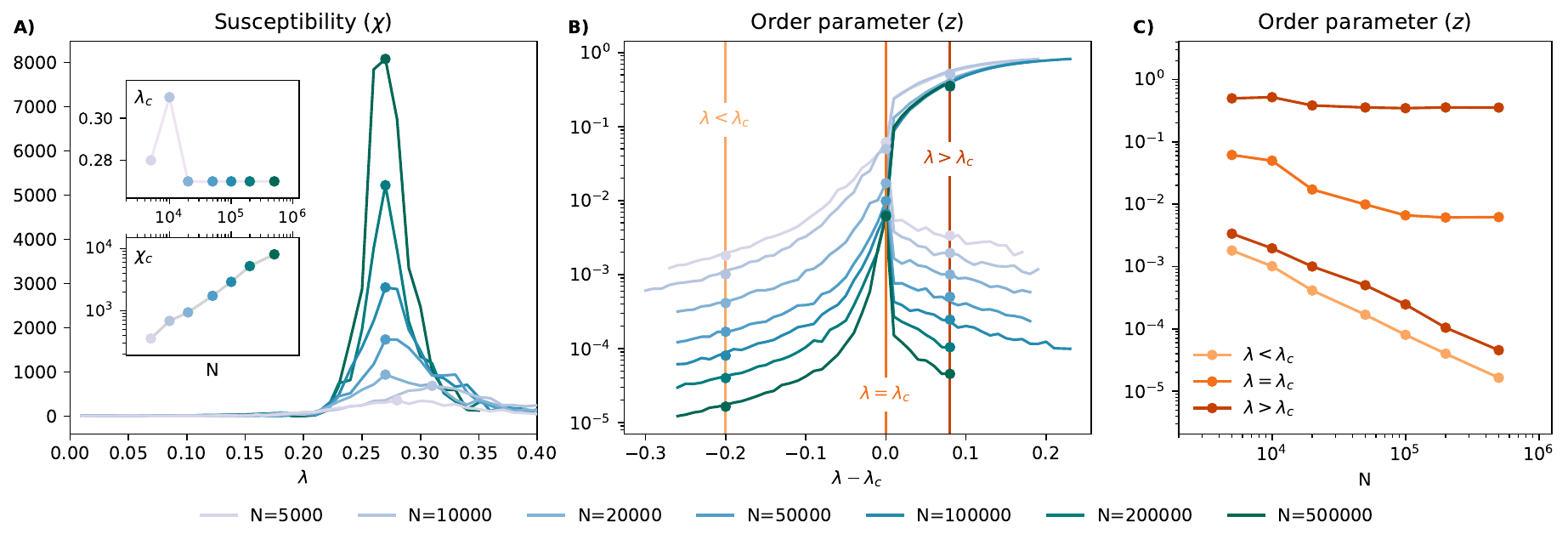}
        \caption{\textbf{Finite-size analysis on a homogeneous hypergraph.} We show the dependence of the susceptibility, A), and the order parameter, B), on the system size.) Using the peak of the susceptibility, we identify the critical point and show that it converges to the value $\lambda_c=0.27 \pm 0.01$. The order parameter is plotted to highlight the subcritical regime, which scales as $N^{-1}$, the critical regime, which also goes to zero but sublinearly, and the supercritical regime. The latter has two branches, one independent of $N$ and one that scales as the subcritical regime. The synthetic hypergraph was generated with a cardinality and degree sequence distributed according to a truncated Poisson for $k \geq 2$, and $|e| \geq 2$ with $\nu = 4.0$, leading to an average of approximately $4.3$. We ran $\Nsim = 10^3$ independent simulations with the following parameters: $\ContThreshold=0.3$, $\AnniThreshold=0.2$, and we used $N^{-\frac{1}{2}}$ as the threshold for computing the susceptibility (see Sec.~\ref{Sec:methods} for details).}
        \label{fig:finite_size}
\end{figure*}

\section{Numerical Analysis}
\label{Sec:numerical}

To study the phase transition of our model, we perform a finite-size analysis for a homogeneous hypergraph. In Fig.~\ref{fig:finite_size}A we show the susceptibility, defined as $\Susceptibility = N \times \left( \langle Z^2 \rangle- \langle Z \rangle^2 \right) \times \langle Z \rangle^{-1}$, as the number of nodes increases. The peak of the susceptibility curve indicates the critical point, which converges to a value around $\lambda_c = 0.27\pm0.01$. In addition, the peak of the susceptibility, $\Susceptibility_{c}$, approximately follows $\Susceptibility_{c} \sim N^{0.68}$ (see the inset in Fig.~\ref{fig:finite_size}A). Complementarily, Fig.~\ref{fig:finite_size}B and~\ref{fig:finite_size}C show the scaling of the order parameters, highlighting the subcritical behavior, $\lambda<\lambda_c$, the critical point, $\lambda=\lambda_c$, and the supercritical behavior, $\lambda>\lambda_c$. The subcritical behavior of the order parameter scales as $z(\lambda < \lambda_c) \sim N^{-1}$ since the size of the rumor is limited to the initial set of infected nodes or a very small fraction of the active nodes. At the critical point, the average fraction of stiflers tends to zero as $\OrderParam(\lambda_c) \sim N^{-0.56}$, which is slower compared to subcritical scaling. Finally, the supercritical regime consists of two branches, one in which $z(\lambda>\lambda_c) \sim O(1)$, the active state, and another that follows $z(\lambda>\lambda_c) \sim N^{-1}$. Note that even if the propagation rate is large enough for a microscopic initial state, there is a non-zero probability that the rumor will die out before reaching a macroscopic state. This is a consequence of the stochastic nature of the process. Note that this probability also decays as $\lambda$ increases.

Fig.~\ref{fig:finite_size}A and B show that the observed phase transitions are continuous. This observation was also confirmed by additional simulations in homogeneous, heterogeneous, and real hypergraphs (see Appendix~\ref{apdx:critical_point_grid} for additional results). Interestingly, this result contrasts with the social contagion model in~\cite{dearruda2020social,FerrazdeArruda2023multistability}, where the transitions are expected to be hybrid~\cite{FerrazdeArruda2023multistability}. Along the same lines, many spreading models in higher-order networks predict discontinuous transitions~\cite{ferrazdearruda2024contagion}. Indeed, in the simplicial contagion~\cite{iacopini2019simplicial}, the quenched mean-field approximation predicts that to have a continuous phase transition, we must also have enough pairwise connections to drive such a transition~\cite{ferrazdeArruda2021phase}. This suggests that the continuous phase transition may be related to the annihilation mechanism, which depends on interaction with other individuals rather than spontaneous transitions, as in the models mentioned above. Strengthening this argument, the authors in~\cite{Landry2020} proposed a higher-order healing mechanism, where both spreading and healing depend on the product of the states of the neighbors. In this case, only continuous transitions were observed.

\begin{figure*}[t]
    \centering
    \includegraphics[width=\linewidth]{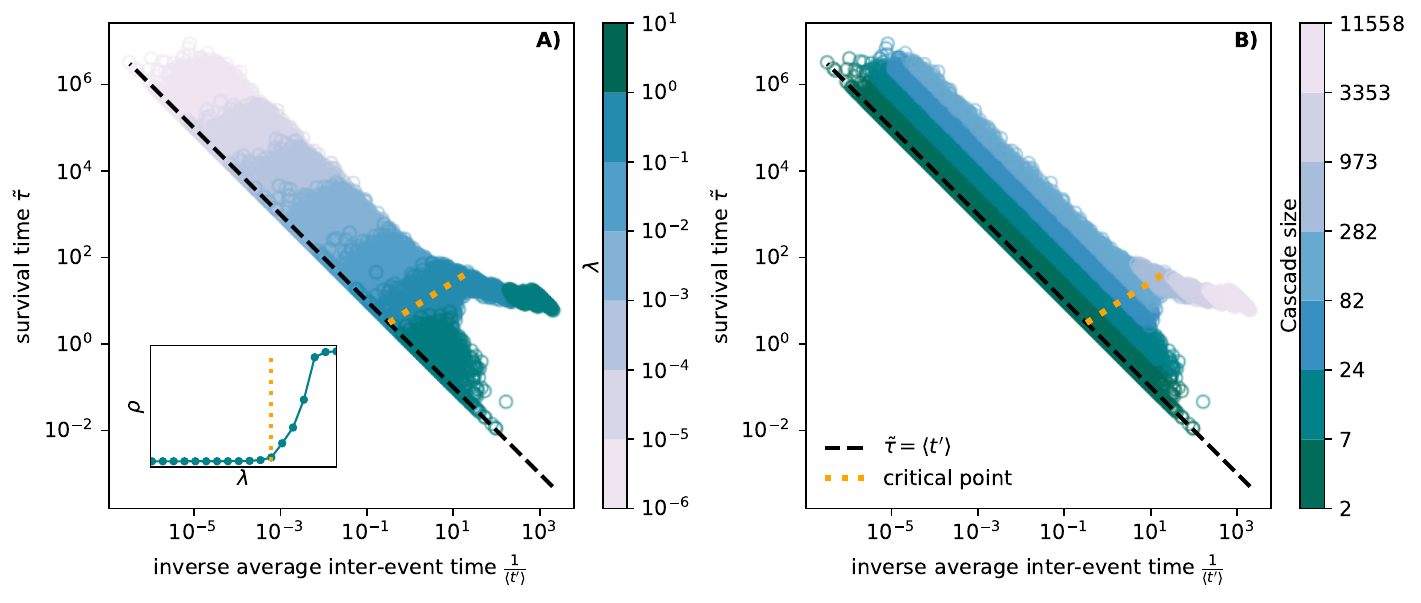}
    \caption{\textbf{Analysis of simulated inter-event time on a heterogeneous hypergraph.} We estimate the average inter-event time and the survival time for the hypergraph rumor model on a heterogeneous hypergraph simulating different contagion rates. In A), we show the different behavior below and above the transition point. We also show the phase transition and the critical point in a small subplot. In B) we show the sizes of the cascades. We consider the time points to be the time at which a contagion process occurred. We discard the annihilation processes from the time series because they are not observable in real scenarios. We considered a hypergraph with $N=10^4$ nodes, and both degree and cardinality follow power law distributions as $P(k) \sim k^{-\gamma}$ and $P(|e|) \sim |e|^{-\gamma}$ with $\gamma = 2.3$. The simulation parameters are $\AnniThreshold = 0.08$, $\ContThreshold = 0.5$, and we run $\Nsim = 10^3$ independent simulations for each value of $\lambda$.}
    \label{fig:interevent_sim}
\end{figure*}

Next, we focus on the analysis of rumor cascades. It is often impossible to obtain the phase diagram of a real social system. Usually, for real information and rumor spreading, we only have the content of the message and its time (e.g., mobile messenger datasets). This can be used as a probe for spreading events. In general, however, it is not possible to know exactly when a person becomes a stifler because this process is inherently related to the person's loss of interest in spreading. Thus, more meaningful observables would be related to cascade sizes and survival times. In Fig.~\ref{fig:interevent_sim} we simulate the rumor model on synthetic heterogeneous hypergraphs. In Fig.~\ref{fig:interevent_sim}A and B, we show how the survival time changes as a function of the average interevent time for different values of $\lambda$ and cascade size, respectively. In this experiment, we use $\lambda \in [10^{-6}, 10^{1}]$, which covers all regimes. We find the transition point to be around $\lambda_c = 0.4$, as highlighted in the inset of Fig.~\ref{fig:interevent_sim}A. The critical point is shown with a dotted line and corresponds to the average of the region where two distinct branches begin to appear. The supercritical regime is the upper branch after the critical point. In this case, the cascades reach a macroscopic fraction of the population. Note that the critical point shows a high variability of cascade sizes, which is expected from second-order phase transitions.

\section{Empirical evidence: The Pushshift Telegram dataset}
\label{Sec:Interevent}

To validate our model, we compare it to real examples of rumor propagation in higher-order systems. Mobile messenger apps allow data collection with timestamps, i.e., it is possible to know exactly the time at which messages are sent, which is equivalent to rumor activity in our problem. This allows us to track each contagion process in time and define a cascade of contagion events. We can also know when the last message on a particular topic was sent and identify that time as the time when the rumor died out. This provides an estimate of the rumor's absorption time.

\begin{figure*}[t]
    \centering
    \includegraphics[width=\linewidth]{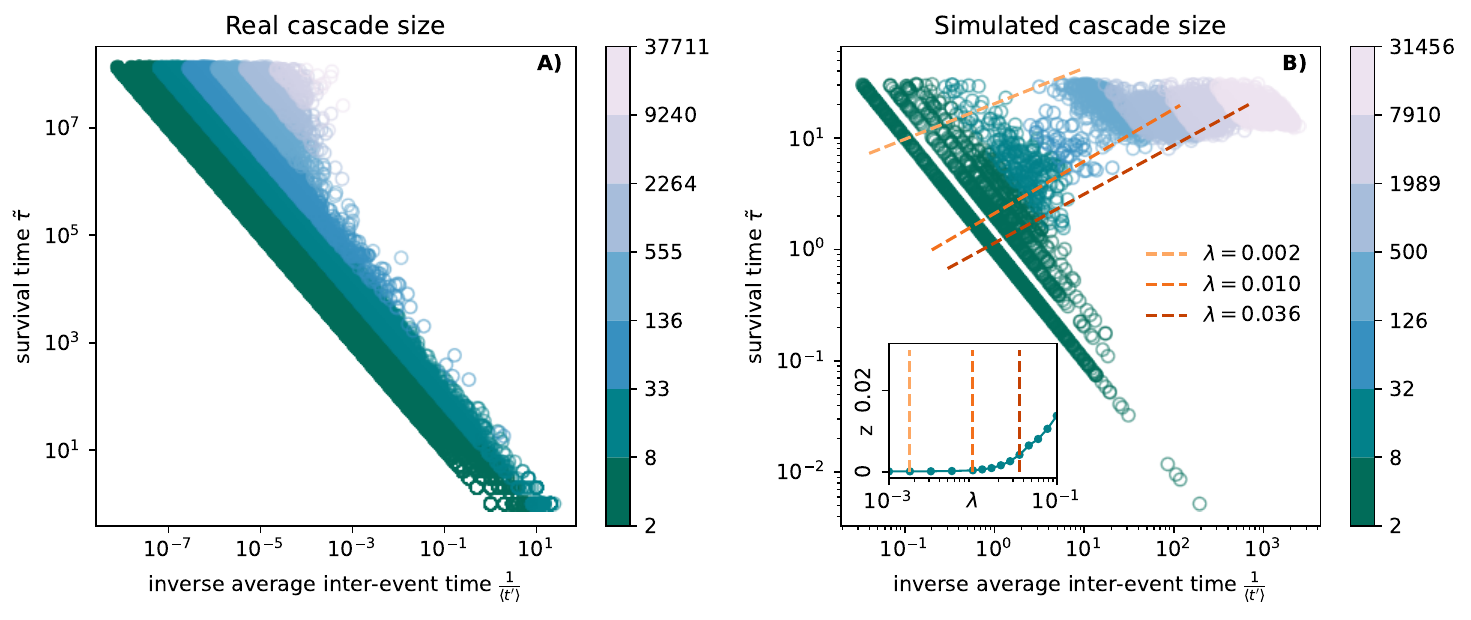}
    \caption{\textbf{Comparison between real observed and simulated cascades.} In A), we plot the inverse of the average inter-event time versus the survival time for the observed cascade of Telegram messages forwarded from one group to another. For this calculation, we only considered messages containing a URL. In the main panel of B), we show simulated cascades on the Telegram hypergraph using the rumor model with parameters $\alpha=1$, $\ContThreshold = 0.1$, $\AnniThreshold = 0.01$. The simulation is plotted with a cutoff on the survival time to replicate the real behavior in which we can observe cascades up to a certain time. The inset of B) features a phase transition from the contagion rate $\lambda$ versus the order parameter $z$. Note that the simulations shown in B) are generated using a range of lambda close to the critical point. Furthermore, the discrepancy in time scale between the real and simulated case is not an issue since the parameter $\alpha$ in the simulations can be adjusted accordingly to rescale the time.}
    \label{fig:telegram}
\end{figure*}

To compare the behavior of our model with real-world rumor dynamics, we analyze the Pushshift Telegram dataset~\cite{baumgartner2020pushshift} (see Appendix~\ref{apdx:dataset} for details). We identify a rumor as messages containing a fixed URL, which allows us to construct a cascade of messages and compute the corresponding average time between events. Additionally, we measure the survival time of a rumor as the time elapsed between the first and last time it was sent. The average inter-event time and survival time for each cascade are shown in Fig.~\ref{fig:telegram}A.

Furthermore, in Fig.~\ref{fig:telegram}B, we show the simulated cascades in the telegram hypergraph. We use a contagion threshold of $\ContThreshold=0.1$ in the simulation based on an estimate derived from the intersection of the Telegram groups (see Appendix~\ref{apdx:dataset}). As it is not possible to estimate the annihilation threshold from the data, we have set $\AnniThreshold=0.01$ to consider a regime where both exponential and power-law decay are present. In particular, we focus on values of $\lambda$ around the critical point since this is the region where the cascades most closely resemble the real case. Note the similarities between the real and simulated cascades shown in Fig.~\ref{fig:telegram}. Given that the model is Markovian, one could rescale the rates in the simulation so that they are in similar ranges to the real data. We chose these rates to avoid numerical problems. It should also be noted that the real case does not have the supercritical branch, while our simulation does. The fact that rumors in the Telegram dataset form cascades that fall into subcritical and critical behavior is in accordance with~\cite{notarmuzi2022universality}. Importantly, Fig.~\ref{fig:telegram} shows that our model captures the cascade behavior in the real data. Thus, the similarities between our simulations and the real data provide additional evidence that rumors spread around criticality in real scenarios, consistent with the findings in~\cite{notarmuzi2022universality} (see Fig.~\ref{fig:telegram}).

Complementarily, Appendix~\ref{apdx:email} shows results using the email dataset~\cite{Benson-email-dataset}. As can be seen, we obtained similar results, confirming our previous findings.

\section{Discussion and Conclusions}\label{Sec:discussion}

\subsection{Group-based annihilation}

We introduced a higher-order rumor model that incorporates group-based annihilation, which follows the original proposal of the Daley-Kendall model~\cite{daley1964epidemics}, i.e., that rumor annihilation occurs through interaction with individuals who are already familiar with the rumor. Additional motivation for our threshold-based annihilation approach can be found in~\cite{Granovetter}. Although the ideas in~\cite{Granovetter} were motivated by contagion, activation was proposed as a process associated with individuals, similar to our approach.

To our knowledge, the first higher-order healing mechanism in the context of higher-order networks was proposed in~\cite{Landry2020}, where the motivation was that if a trend is popular in groups, then this might make individuals less likely to adopt such a trend. They also called this the ``hipster effect.'' Although this mechanism is similar to annihilation in rumor spreading, the model proposed in~\cite{Landry2020} still has a single absorbing state and is intended to be an extension of the SIS model.

In addition, as future work, exploring alternative versions of higher-order annihilation may be useful in other rumor-spreading contexts, especially when the cardinality of the hyperedge is small, and it becomes more reasonable for a spreader to track the state of individual nodes rather than entire groups. An interesting open question, then, is how different annihilation processes shape the dynamics and whether these models can be unified into the same class.

\subsection{Subcritical behavior}

Our model has two different subcritical behaviors, exponential or power-law decay to an absorbing state. Interestingly, the exponential decay resembles the behavior of an SIR model in a graph~\cite{Mieghem2009,deArruda2028}, while the power-law decay has been observed in the Maki-Thompson rumor model~\cite{ferrazdearruda2022subcritical}. The main mechanism behind this variety of behavior is the threshold-based annihilation process, which is higher-order by definition. If the threshold is low enough, the subcritical behavior is exponential because the initial condition is sufficient to trigger its own annihilation process, similar to a spontaneous process. Note that this happens regardless of the propagation rate. On the other hand, if the threshold is moderately high and $\lambda \ll \alpha$, at least one spread must occur for the annihilation threshold to be reached. This produces a power-law behavior of the form $\tau \propto \frac{1}{\lambda}$ similar to the subcritical behavior of the Maki-Thompson model.

In heterogeneous systems, which is the case for many real systems, both behaviors can be observed depending on the initial condition. In fact, the parameter space in Fig.~\ref{fig:diagram}A changes slightly in the presence of heterogeneity. In particular, for a power-law hypergraph with $P(k) \sim k^{-2.3}$ and $P(|e|) \sim |e|^{-2.3}$ we observed that the ``always supercritical region becomes larger while the power-law decay region becomes smaller (see Appendix~\ref{apdx:critical_point_grid} for more details).

The analysis of the subcritical behavior of our model is particularly interesting because the Maki-Thompson and SIR models have traditionally been studied separately. By incorporating these behaviors into a single framework, our model opens up new avenues of investigation. Future work can focus on rigorously calibrating the model to real-world scenarios and determining whether a dominant annihilation process emerges or whether a mixed effect occurs naturally.

\subsection{Critical behavior}

It is common to observe discontinuous phase transitions in dynamical processes in higher order networks~\cite{ferrazdearruda2024contagion}. In our experiments, however, we observed only continuous phase transitions. The main differences between our model and the social contagion counterpart proposed in~\cite{dearruda2020social} are the number of absorbing states and the annihilation mechanism. Since the SIR dynamics on simplicial complexes also presents discontinuous transitions~\cite{malizia2025}, this suggests that the number of absorbing states is not sufficient to explain the nature of our transition. However, in~\cite{Landry2020}, higher-order healing was studied where discontinuous transitions were not found. Thus, it provides some evidence that the continuous phase transitions of our model are related to non-spontaneous annihilation. Along similar lines, another relevant question would be whether it is possible to have discontinuous transitions in our model, either hybrid as in~\cite{dearruda2020social, FerrazdeArruda2023multistability} or even first-order. In this case, what would be the sufficient and necessary conditions for such behavior? The formal verification of these questions is an open problem.

Despite the nature of the transition, the critical point changes as a function of both the structure and the contagion and annihilation thresholds. The interplay between these quantities is not trivial. Indeed, we find that the region of the parameter space $\Theta_\lambda \times \Theta_\alpha$ in which a phase transition is observed is significantly smaller for power-law hypergraphs than for the truncated Poisson hypergraphs (see Appendix~\ref{apdx:critical_point_grid} for additional experiments). Understanding the interplay between the structure and the contagion and annihilation thresholds is another open problem.

\subsection{Rumor spreading occurs near criticality}

Our results are consistent with the existing literature and provide evidence that real-world rumor spreading tends to occur near criticality~\cite{notarmuzi2022universality}. By comparing message cascades in Telegram with our simulations, we found that the contagion rate at which the cascades align is close to criticality. We repeated the experiment with the email dataset~\cite{Benson-email-dataset} and obtained similar results (see Appendix~\ref{apdx:email}). Moreover, our model may provide a mechanistic explanation for the cascades observed in the real data, as it is able to reproduce some of their main features. To further validate this hypothesis, more real-world examples of rumor spreading and controlled experiments in higher-order systems need to be considered.

\section{Methods}
\label{Sec:methods}

\subsection{Gillespie algorithm}

We simulate the rumor model using continuous-time Monte Carlo simulations via the Gillespie algorithm~\cite{gillespie1977stochastic}. At each step, we track all contagion and annihilation processes that meet their respective thresholds and trigger them according to random samples from exponential distributions. Each process is a Poisson process parameterized by $\lambda$ (contagion) or $\alpha$ (annihilation). The threshold parameters $\ContThreshold, \AnniThreshold \in (0,1)$ are evaluated in each process by rounding their product with the hyperedge cardinality (for contagion) or node degree (for annihilation). When a process is triggered, it updates node states and may abort pending processes (e.g., nodes transitioning to stiflers reduce the contagion threshold count).

The algorithm is initialized on a hypergraph where all nodes are ignorant except those in a single randomly chosen hyperedge, which are initialized as spreaders. The algorithm terminates when no more transitions are possible, typically when all spreaders become stiflers. However, in some cases, due to the annihilation threshold condition, a few spreaders may remain active at the end.

To characterize the transition, we denote as $Z^s$ the total number of stiflers at the end of simulation $s$, then the order parameter $\OrderParam$ is given by
\begin{equation}
   \OrderParam = \langle Z \rangle = \frac{1}{N} \left(\frac{1}{\Nsim}\sum_s^{\Nsim} Z^s \right),
\end{equation}
where $\langle \cdot \rangle$ is the expectation operator, $\Nsim$ is the number of simulations. Then, we can define the susceptibility as
\begin{equation}
    \Susceptibility = N \frac{\langle Z^2 \rangle- \langle Z \rangle^2}{\langle Z \rangle},
\end{equation}
and we find the critical point $\lambda_c$ as
\begin{equation}
    \lambda_c = \operatorname*{argmax}_\lambda \Susceptibility .
\end{equation}

Even above the critical point, some simulations may end with only a small number of stiflers due to stochastic fluctuations and the presence of thresholds. This effectively creates two branches: the upper critical branch and an inactive branch caused by randomness. If both branches were included, the calculated susceptibility would be significantly overestimated. To mitigate this problem, when calculating the susceptibility and the order parameter, we only consider simulations where the final number of stiflers satisfies $Z/N > N^{-\frac{1}{2}}$. If no simulations fulfill this criterion, the system is classified as subcritical, and all simulation calculations are included.

\subsection{Synthetic hypergraphs}

We simulate our model on synthetic hypergraphs with truncated Poisson and power-law degree and cardinality distributions. To generate the hypergraphs, we used an algorithm based on three steps: (i) an unrestricted matching, (ii) a brute-force fixing algorithm that swaps repeated nodes on the hyperedges, and (iii) a random swap step (using the swap proposed in~\cite{Chodrow2020}) that ensures that the final hypergraph is uniformly sampled from the space of possible hypergraphs. We perform $10^4$ swaps for each hypergraph. This algorithm has been proposed and systematically tested in~\cite{Arruda2024} and used to generate hypergraphs in~\cite{rw_hypergraphs}. This algorithm can generate hypergraphs with the desired number of nodes, cardinality, and degree distribution, and it can also fix the maximum and minimum degree and cardinality. The distributions are truncated and we choose the minimum degree $k_{\text{min}} = 2$ and the maximum degree $k_{\text{max}} =\sqrt N$, while the minimum cardinality $|e|_{\text{min}} = 2$ and the maximum cardinality $|e|_{\text{max}} = \sqrt N$.

\subsection{Interevent time analysis}

Given a  cascade $\{t_0, t_1, \dots ,t_s\}$ of size $s+1$, we measure the survival time $\tilde{\tau}$ as
\begin{equation}
    \tilde{\tau} = t_s -t_0,
\end{equation}
and average interevent time $\langle  t'\rangle$ as
\begin{equation}
    \langle  t'\rangle = \frac{1}{s}\sum_{i=0}^{s-1} (t_{i+1} - t_{i}) = \frac{t_{s}- t_0}{s} = \frac{\tilde{\tau}}{s}.
\end{equation}

\subsection{Derivation of threshold bounds}

In Sec.~\ref{Sec:The model}, we introduced three bounds that separate the regimes shown in Fig.~\ref{fig:diagram}A. The first bound, $\AnniThreshold^1$, represents the threshold below which every node in the hypergraph recovers exponentially fast, provided the rumor spreads. In this regime, a spreader node becomes a stifler as soon as it belongs to a single active hyperedge. Mathematically, this corresponds to the condition where the term $Y_i H(S_i-k_i\AnniThreshold)$ in Eq.~\ref{eq:model_dy} is equal to $1$ for all $i$. In order to recover, $Y_i = 1$, implying that at least one hyperedge is active. Then $S_i\geq1$, leading to the condition that if
\begin{equation}
    \AnniThreshold \leq \AnniThreshold^1 = \frac{1}{k_{\text{max}}},
\end{equation}
where $k_{\text{max}}$ is the maximum degree, all nodes recover exponentially fast.  In this regime, the model behaves similarly to a SIR model with critical mass contagion.

Conversely, $\AnniThreshold^2$ is the value above which the rumor always reaches a finite fraction of the population, even for an arbitrarily small but positive $\lambda$.  Estimating this bound is more complex since it also depends on $\ContThreshold$ (see Appendix~\ref{apdx:critical_point_grid} for an example).  However, we can give an upper bound as
\begin{equation}
\AnniThreshold>\AnniThreshold^2 = \frac{1}{k_{\text{min}}},
\end{equation}
where $k_{\text{min}}$ is the minimum degree.

Finally, there is an inactive region above the value $\ContThreshold^1$, where the rumor never reaches a finite fraction of the population.  
This happens when the term $X_i H(T_j-|e_j|\ContThreshold)$ in Eq.~\ref{eq:model_dy} is always $0$ for each hyperedge $e_j$. To be infected, $X_i = 1$, implying that the sum of infected nodes in a hyperedge $e_j$ is bounded by $T_j< |e_j|-1$. Thus, any $\ContThreshold > (|e_j|-1)/|e_j|$ makes the contagion of hyperedge $e_j$ impossible.
In general, no contagion is possible if
\begin{equation}
    \ContThreshold>\ContThreshold^1 = 1 - \frac{1}{|e|_{\text{max}}},
\end{equation}
where $|e|_{\text{max}}$ is the largest cardinality. However, this bound tends to overestimate the true threshold.  For example, in Appendix~\ref{apdx:critical_point_grid}, we show an example where $|e|_{max} = 10$, but there is no transition when $\ContThreshold>0.75$.

Note that these bounds are sharp for uniform regular hypergraphs. However, in the presence of heterogeneity, the boundaries of the different subcritical behaviors may be a function of the contagion and annihilation thresholds.  For an example, see figures~\ref{supfig:grid} and~\ref{supfig:grid_Poisson}, where heterogeneity bends the boundary of the always active region.

\begin{acknowledgments}
    K.A.O~ was partially supported by the European Union through ERC grant (ID-COMPRESSION, grant number: 101124175). Views and opinions expressed are however those of the author(s) only and do not necessarily reflect those of the European Union or the European Research Council Executive Agency. Neither the European Union nor the granting authority can be held responsible for them.
    P.T.~ and Y.M.~ were partially supported by the Government of Arag\'on, Spain and ``ERDF A way of making Europe'' through grant E36-23R (FENOL), and by Ministerio de Ciencia e Innovaci\'on, Agencia Espa\~nola de Investigaci\'on (MCIN/AEI/ 10.13039/501100011033) Grant No. PID2023-149409NB-I00. We acknowledge the use of the computational resources of COSNET Lab at Institute BIFI, which were funded by Banco Santander (grant Santander-UZ 2020/0274) and the Government of Arag\'on (grant UZ-164255). The funders had no role in study design, data collection and analysis, decision to publish, or preparation of the manuscript.
    GFA was partially supported by the São Paulo Research Foundation (FAPESP), Process Number 2024/16711-8.
\end{acknowledgments}

\appendix

\section{Dataset}
\label{apdx:dataset}

We describe here some properties of the Pushshift Telegram hypergraph~\cite{baumgartner2020pushshift} and Email hypergraph~\cite{Benson-email-dataset}.

\subsection{Pushshift Telegram dataset}

We studied the rumor propagation model using the Pushshift telegram dataset~\cite{baumgartner2020pushshift}. This dataset contains $317,224,715$ messages from $2,200,040$ users. First, we build a hypergraph that maps each public channel and/or group identifier (ID) listed in each message's metadata to a hyperedge and each user ID to a node.

Since about $8.1\%$ of the more than $317$ million messages are forwarded from group to group, we use this information, whenever available, to assign the hyperedge membership of the user who posted the message to both groups. This results in a first partial reconstruction of the telegram hypergraph with $141,824$ nodes and $143,343$ hyperedges.

We then do the same with the rest of the messages, which only contain the ID of the group that received the message and attribute the one-way hyperedge membership to the posting user. Adding this new information to the existing hypergraph results in a total of $2,200,040$ nodes and $143,501$ hyperedges. Finally, we keep only the largest connected component of this hypergraph, which comprises almost all of its size. The resulting hypergraph is examined in section \ref{Sec:Interevent}, which has $2,199,885$ nodes and $143,475$ hyperedges.

\begin{figure}[t]
    \centering
    \includegraphics[width=\linewidth]{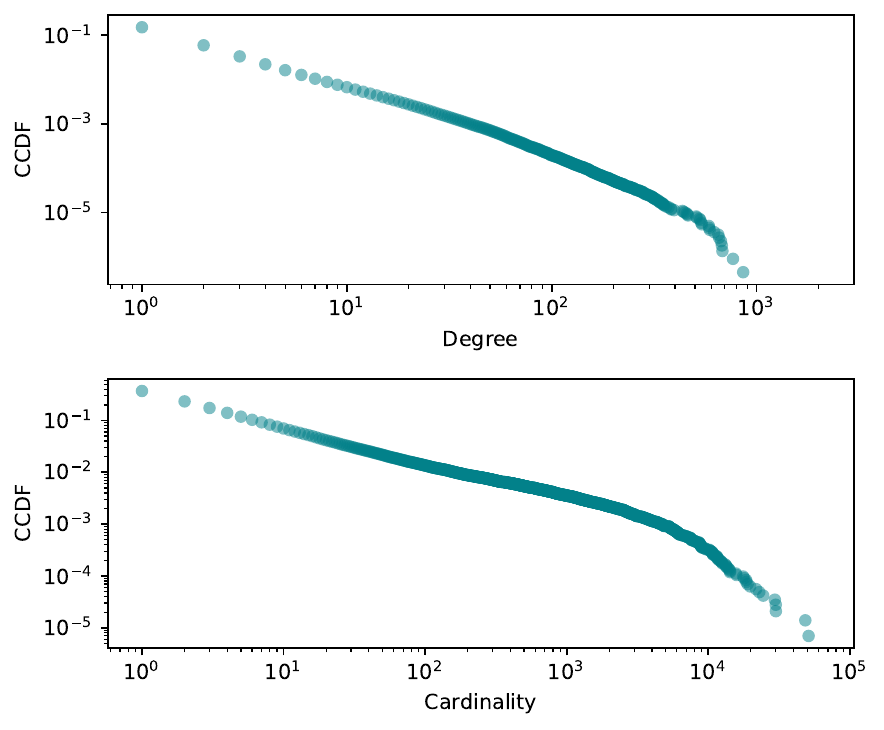}
    \caption{\textbf{The Complementary Cumulative Distribution Function of the Pushshift Telegram Hypergraph.} We show the complementary cumulative distribution function (CCDF) of the cardinality and degree distributions. Both distributions are heterogeneous.}
    \label{supfig:Telegram_CCDF}
\end{figure}

As shown in Fig.~\ref{supfig:Telegram_CCDF}, the hypergraph has heavy-tailed distributions for both cardinalities and degrees. The average cardinality is $22.14$, but the $100$ largest groups have at least $6,088$ members (up to a maximum of $61,377$). In terms of degrees, users are connected to $1.44$ groups on average, and the $100$ largest degrees range from $217$ to $2,035$. However, groups are able to connect many users simultaneously, which means that each user has $12,057$ neighbors on average.

Importantly, the message IDs present in the metadata do not uniquely identify messages across the dataset but only within each group. For this reason, we build cascades of information sharing across groups by tracking Uniform Resource Locators (URLs) instead since we can then say that it is the same rumor that is spreading throughout the system. Sent messages that contain the same URL are assumed to be about the same rumor. There are $2,314,550$ unique URLs that appear in at least two messages in the dataset. Once a URL is fixed, we define a cascade as the ordered time series containing the time when messages containing that URL were sent from one group to another.

\begin{figure}[t]
    \centering
    \includegraphics[width=\linewidth]{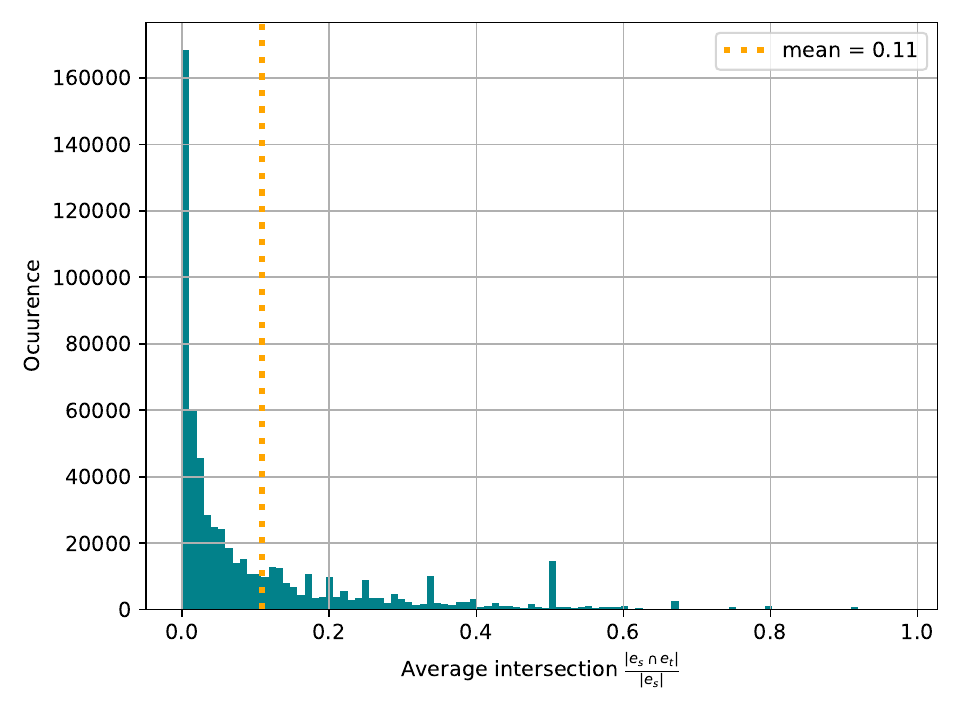}
    \caption{\textbf{Estimating $\ContThreshold$ from the intersection of real active hyperedges.} We show the distribution of the size of the intersections between sending and receiving groups containing a given URL, normalized by the size of the sending group. The mean value of this distribution is denoted by the orange dashed line.}
    \label{supfig:estimate_theta}
\end{figure}

The data allows us to roughly estimate the value of the contagion threshold as $\ContThreshold \approx0.1$. To do this, we select all messages containing a URL for which we have information about the sending and receiving groups and then compute the intersection of the two groups, normalized by the size of the sending group. This estimation is based on the idea that for one hyperedge to trigger another, it must share as many nodes with the next hyperedge as required by $\ContThreshold$. The results are shown in Fig.\ref{supfig:estimate_theta}.

\subsection{Email dataset}

The email dataset was taken from~\cite{Benson-email-dataset} already in the form of a hypergraph. We use the complete data without any restriction on the size of the hyperedges. There are $1,005$ nodes and $235,263$ hyperedges, of which $25,919$ are unique. For the simulation shown in Appendix~\ref{apdx:email}, we used the hypergraph composed of unique hyperedges. Specifically, we had information about the nodes (senders and receivers of the emails) in the hyperedges (the emails) and at what time these hyperedges were created. However, unlike the telegram dataset, we don't have any metadata about the emails. We don't know the subject or body of the emails, if an email is a reply, or if the email was forwarded to others. The lack of metadata makes it difficult to build the cascades of emails, and we have to make assumptions. We consider a cascade of emails to be a time series of emails sent between the same group of people. This is a rough estimate, since we are missing emails that are forwarded to a person outside the original group of people involved, or we can consider emails on different topics as a single rumor because they are exchanged between the same group of people.

\begin{figure}[t]
    \centering
    \includegraphics[width = \linewidth]{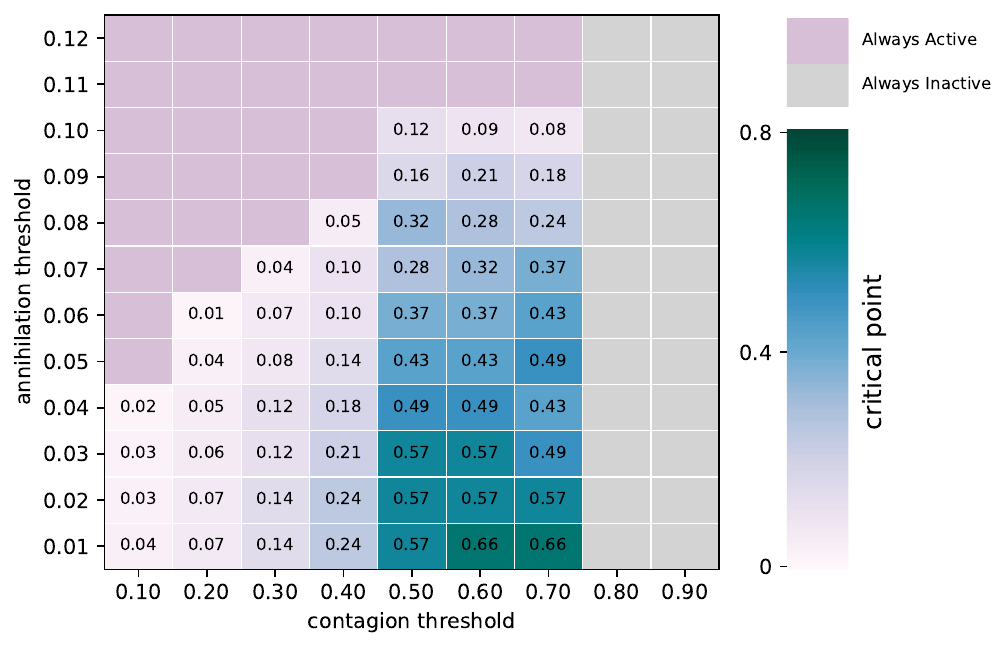}
    \caption{\textbf{Critical points with different thresholds (heterogeneous hypergraph).} We simulate the rumored model on a heterogeneous hypergraph with $N = 10^4$ nodes and power-law degree and cardinality distributions (for both, the power-law exponent is $2.3$) for different annihilation and contagion threshold values. The critical point was calculated as the value of lambda for which the susceptibility of the number of stiflers is maximal. To avoid a smeared susceptibility, we did not include in the variance calculation all simulations for which the number of stiflers was less than $1/\sqrt N = 0.01$ unless the simulation was in the subcritical regime.}
    \label{supfig:grid}
\end{figure}

\begin{figure}[t]
    \centering
    \includegraphics[width=\linewidth]{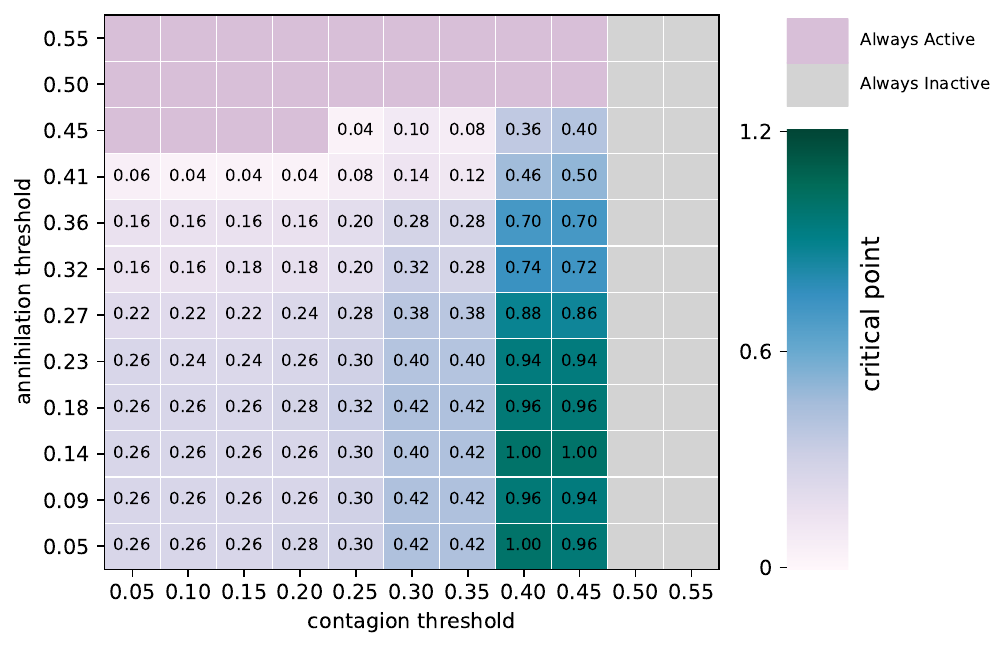}
    \caption{\textbf{Critical points with different thresholds (homogeneous hypergraph).} We simulate the rumor model on a homogeneous hypergraph with $N = 10^4$ nodes and truncated Poisson degree and cardinality distributions (for both, $\nu = 2$, and the average is approximately $2.9$) for different values of the annihilation and contagion thresholds. The critical point was calculated as the value of lambda for which the susceptibility of the number of stiflers is maximal. To avoid a smeared susceptibility, we did not include in the variance calculation all simulations for which the number of stiflers was less than $1/\sqrt N = 0.01$ unless the simulation was in the subcritical regime.}
    \label{supfig:grid_Poisson}
\end{figure}

\section{Critical Points with different thresholds}
\label{apdx:critical_point_grid}

We characterize how the transition point varies with the choice of annihilation and contagion thresholds. We show in Fig.~\ref{supfig:grid} the case of a heterogeneous hypergraph with $N = 10^4$ nodes and with both cardinality and degree sequence distributed as a truncated power law with exponent $\gamma = 2.3$ and maximum degree or cardinality $\sqrt{N} = 10^2$. As a complement, in Fig.~\ref{supfig:grid_Poisson}, we show the case of a homogeneous hypergraph with $N = 10^4$, where both the cardinality and degree distributions follow a Poisson distribution with mean $2.0$. The reader may recognize the pattern shown in Fig.~\ref{fig:diagram} of the main text.

\begin{figure*}[t]
    \centering
    \includegraphics[width=\linewidth]{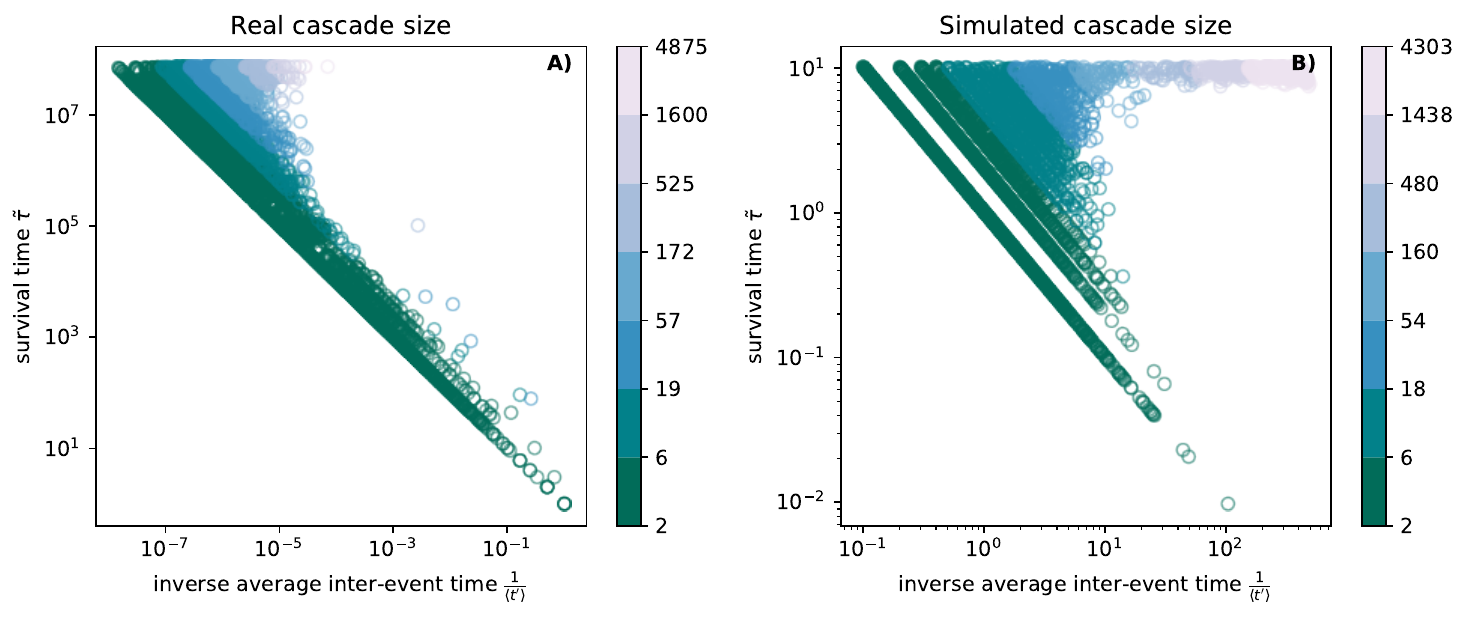}
    \caption{\textbf{Comparison between the real observed and simulated cascades.} In A), we plot the inverse of the average inter-event time versus the survival time for the observed cascade of emails sent between the same people. The simulation is plotted with a cutoff on the survival time to replicate the real behavior where we can observe cascades up to a certain time. Note that the simulations plotted in B) are generated using a range of lambda close to the critical point. Furthermore, the discrepancy in time scale between the real and simulated case is not an issue since the parameter $\alpha$ in the simulations can be adjusted accordingly to rescale the time.}
    \label{supfig:email}
\end{figure*}

\section{Real hypergraph: E-mail}
\label{apdx:email}

To further support our findings, we report additional results for the email dataset. In Fig.~\ref{supfig:email}, we perform the same inter-event time analysis as for the telegram dataset. Also, the similarity between the simulated and observed data is remarkable. However, it is important to note that the construction of rumor cascades differs between the two datasets. In the case of emails, we consider a cascade to be a time series of emails sent between the same individuals, which means that the rumor remains confined within the same hyperedge. Due to the lack of information about email content or type, it necessarily omits forwarded emails and alternative propagation mechanisms.

\end{document}